\documentstyle[aps,preprint,prl]{revtex}

\draft

\begin{document}
\title{Possible new vortex matter phases in 
 Bi$_2$Sr$_2$CaCu$_2$O$_8$}
\author{D. T. Fuchs$^{1}$, E. Zeldov$^{1}$, 
 T.~Tamegai$^{2}$, S.~Ooi$^{2}$, M.
 Rappaport$^{3}$, and H. Shtrikman$^{1}$}
\address{$^{1}$Department of Condensed Matter Physics,
 The Weizmann Institute of Science, Rehovot 76100, Israel}
\address{$^{2}$Department of Applied Physics, The University of
 Tokyo, Hongo, Bunkyo-ku, Tokyo, 113, Japan}
\address{$^{3}$Physics Services,
 The Weizmann Institute of Science, Rehovot 76100, Israel}
\date{\today}
\maketitle

\begin{abstract}
The vortex matter phase diagram of Bi$_2$Sr$_2$CaCu$_2$O$_8$ 
crystals is analyzed by investigating vortex penetration through the 
surface barrier in the presence of a transport current. The strength 
of the effective surface barrier and its nonlinearity and asymmetry 
are used to identify a possible new ordered phase above the 
first-order transition. This technique also allows sensitive 
determination of the depinning temperature. The solid phase below 
the first-order transition is apparently subdivided into two phases 
by a vertical line extending from the multicritical point.  
\end{abstract}



\newpage

The traditional view of type II superconductors was that an ordered 
solid vortex lattice exists over the entire mixed state phase 
diagram. For high-temperature superconductors, in contrast, it has 
been suggested theoretically \cite{nelson,review}, and demonstrated 
experimentally \cite {safar,kwok,melt,schilling}, that in clean 
samples the vortex lattice undergoes a first-order melting 
transition and thus two phases, vortex solid and vortex liquid, are 
formed. Recent investigations \cite 
{borisprl,boris-disorder,cubitt,lee}, however, have shown that the 
vortex matter phase diagram is comprised of three distinct phases: a 
rather ordered lattice at low fields, a highly disordered solid at 
high fields and low temperatures, and a vortex fluid phase at high 
temperatures and fields. Such a phase diagram is supported by a 
number of recent theoretical studies \cite {ertasnelson,gl-vinokur}. 
In this letter we show that the phase diagram is even more complex 
and apparently displays additional new vortex matter phases.

One of the numerous open questions is the structure of the vortex 
fluid phase, and whether it is further subdivided into two phases: a 
liquid of vortex lines, and a decoupled gas of vortex pancakes in 
the CuO planes \cite {glazkosh-Daemen,samoilov,sublimation}. Both of 
these fluid phases are weakly pinned, and therefore display high 
resistivity and reversible magnetization, and thus the existence of 
a possible transition between them is difficult to resolve. We 
demonstrate here a new approach for probing such phase transitions. 
Instead of investigating vortex interaction with the bulk pinning 
potential, we probe vortex penetration through the surface barrier 
(SB).  It has been shown theoretically that the effective height of 
the Bean-Livingston SB, and its non-linear dependence on the driving 
force, depend on whether the vortices are in a solid, liquid, or 
gaseous state \cite{burlachkov,koshelev}. The recently developed 
technique \cite{curdisnature} for measuring the distribution of the 
transport current across the sample is well suited for investigation 
of vortex activation over the SB in the presence of a driving force. 
It also provides valuable information on the bulk transport 
properties at low fields and temperatures at which the sample 
resistivity is below the resolution of conventional transport 
measurements.

Bi$_{2}$Sr$_{2}$CaCu$_{2}$O$_{8}$ (BSCCO) crystals ($T_{c}$ $\simeq 
$ 88K) were grown using the traveling solvent floating zone method 
\cite{motohira}.  Silver/gold contacts were evaporated onto freshly 
cleaved surfaces of single crystals of typical size 150 $\times $ 
1500 $\times $ 10 $\mu $m$^{3}$. The crystals were positioned on an 
array of seven Hall sensors that measure the perpendicular component 
of the magnetic field \cite{melt,curdisnature} as shown in the inset 
to Fig. 1. Transport current of 0.5 to 10mA rms (30 - 400 Hz) was 
applied to the crystal. The resulting self-induced ac field $B_{ac}$ 
was measured by the sensors using a lock-in amplifier 
\cite{curdisnature}. A constant dc magnetic field of 20G to 5000G 
was applied parallel to the c-axis. A total of nine crystals were 
investigated. The characteristic features, which are described below 
for two typical crystals, were observed in all the samples. 

Figure 1d shows the self-field $B_{ac}$ as a function of temperature 
at a dc field of 1000G. There are three main forms of transport 
current flow across the width of the sample, depending on the 
dynamics of vortices \cite {curdisnature}, as indicated 
schematically in Figs. 1a-1c. ({\it 1a}) Vortices are mobile and 
their drift velocity is determined by bulk viscosity or pinning. In 
this case the current flows uniformly across the sample, as it does 
in a normal metal, and $B_{ac}$ decreases monotonically from one 
edge of the sample to the other (Biot-Savart law). Such flow is 
observed at $T>T_{sb}$ in Fig. 1d. ({\it 1b}) Vortices are mobile, 
but their flow rate across the sample is determined by the low 
hopping rate over the SB. In this situation most of the current 
flows at the two edges, where vortices enter and exit the 
superconductor, in order to provide the necessary driving force to 
overcome the barrier. Only a very small fraction of the current 
flows in the bulk to maintain the relatively easy drift of the 
vortices across the bulk. The resulting self-field distribution 
inside the sample (sensors 2~-~7) is opposite both in sign and in 
slope to that for the uniform flow case. This unusual situation 
\cite{curdisnature} is observed over a wide range of temperatures 
$T_{d}<T<T_{sb}$ in Fig. 1d.  The crossover between the SB and 
uniform flow occurs gradually over the interval of 60-80K. At the 
intersection point $T_{sb}$, half of the current flows in the bulk 
and half at the two edges of the sample \cite{curdisnature}. The 
dashed line in Fig. 2 shows the field dependence of $T_{sb}$. ({\it 
1c}) Vortices are immobile for the given applied current. In this 
case the transport current distribution has a characteristic 
Meissner form resulting in $B_{ac}=0$ within the sample. Such bulk 
vortex pinning is observed for $T<T_{d}$. The field dependence of 
the depinning temperature $T_{d}$ is shown in Fig. 2.

Figure 1e presents similar data at a lower field of 550G. At fields 
below 750G another characteristic temperature is present, $T_{FOT}$, 
at which the first-order phase transition (FOT) occurs \cite{melt}. 
Above $T_{FOT}$ practically all the current flows at the edges of 
the sample. At the transition the relative share of the current in 
the bulk is suddenly increased, as seen by the drop in the negative 
self-field signal (linear combination of Fig. 1b with a small 
contribution of 1a). This indicates that the impedance for vortex 
flow in the bulk is enhanced, as expected for a liquid-solid 
transition \cite{review}. However, it is important to note that 
below the transition ({\it i}) the vortices are still mobile for all 
our applied currents, since $|B_{ac}|>0$ within the sample, and 
({\it ii}) the vortex flow rate is still governed by the SB, since 
$B_{ac}$ profile is inverted (with respect to the uniform flow). 
Vortices become immobile due to bulk pinning at the significantly 
lower temperature $T_{d}$. The field dependencies of $T_{d}$ and 
$T_{FOT}$ are shown in Fig. 2. The FOT in this crystal was measured 
independently by detecting the step in the equilibrium magnetization 
\cite{melt} as indicated by the inverted triangles.  Figure 2 also 
shows the position of the second magnetization peak line $B_{sp}$ 
obtained on the same crystal using local magnetization measurements 
\cite {borisprl,boris-disorder}. We emphasize that the observation 
of the $T_{d}$ line below $B_{sp}$ is a result of the much higher 
sensitivity of our technique as compared to transport measurements 
for which the resistivity below FOT is immeasurably low.

At the FOT, a sharp drop in sample resistance by typically two 
orders of magnitude is observed 
\cite{sublimation,simultaneous-watauchi}. This drop is usually 
ascribed to the onset of pinning upon freezing into the solid phase 
\cite{safar,kwok}. The data in Fig. 1e show that in BSCCO the 
situation is more complicated; both above and below the FOT vortices 
are mobile and most of the current flows at the sample edges. The 
drop in measured resistance therefore reflects a sharp increase in 
the height of the SB. So upon freezing, both the bulk pinning and 
the SB increase abruptly, but the vortex flow is still dominated by 
the hopping rate over the SB.

Above the FOT, bulk pinning is very weak and thus is not efficacious 
as a probe of the state of the vortex matter. Activation over the 
SB, on the other hand, is predicted to differ significantly between 
the possible vortex phases \cite{burlachkov,koshelev}. In 
particular, the effective height of the barrier increases as the 
order and the c-axis coherence of the vortex matter increase. For a 
decoupled gas, for example, the pancakes are individually thermally 
activated over the barrier and the effective barrier height is low 
\cite{burlachkov}. For a liquid of vortex lines, in contrast, a 
nucleation loop of a minimum critical size has to be created to 
allow vortex penetration, resulting in a higher effective barrier 
\cite{burlachkov}. In a solid phase, the lattice has to deform in 
order to accommodate the penetrating vortex, which further increases 
the barrier height \cite{koshelev}. Thus for various processes, the 
effective barrier height and its non-linear current dependence are 
different.

Another important property of the SB is the asymmetry between vortex 
entry and exit. At equilibrium magnetization, the heights of the 
barriers for vortex entry and exit are identical \cite{Clem}. As a 
result, in the limit of low transport current in the presence of 
thermal excitations, the transmission through the barrier is linear 
with current, and hence the current is equally divided between the 
two edges of the sample (Fig. 1b).  However, due to its asymmetric 
structure, the barrier height for vortex exit decreases faster with 
increasing the current as compared to vortex entry 
\cite{Clem,burlachkov}. As a result, at high currents the division 
between the two edges is unequal, with a larger fraction of the 
current flowing at the vortex entry edge. For an ac transport 
current, on the positive half cycle, vortices enter at the left edge 
and exit at the right (see inset of Fig. 1), and vice versa on the 
negative half cycle. Accordingly, the current on the left edge will 
be larger on the positive half cycle and smaller on the negative, 
resulting in a strong second harmonic signal $B_{ac}^{2f}$ (Fig. 
3a). This is in sharp contrast to bulk pinning which gives rise only 
to odd harmonics due to the fact that the bulk I-V characteristics, 
though nonlinear, are symmetric with respect to the current 
direction \cite{review}. The local I-V of SB, in contrast, is 
asymmetric resulting in even harmonics.  The asymmetry between the 
edges grows with the current, and hence for a given barrier strength 
$B_{ac}^{2f}$ is low at small ac currents and high at large 
currents. Equivalently, for a given ac current a strong barrier 
results in a low second harmonic whereas a weak barrier gives rise 
to a large $B_{ac}^{2f} $. We use this effect to analyze the 
activation over the SB as follows.  Figure 3a shows $B_{ac}^{2f}$ at 
10 mA ac current at 300G and 1000G dc field.  For 300G a large 
$B_{ac}^{2f}$ is observed at elevated temperatures indicating a 
relatively weak SB. At $T_{FOT}$ a discontinuous drop in 
$B_{ac}^{2f}$ is observed below which the signal vanishes rapidly. 
For $T_{d}<T<$ $T_{FOT}$ vortex dynamics is dominated by the SB and 
most of the current flows on the edges, yet $B_{ac}^{2f}$ is 
vanishingly small. As discussed above, this shows that the effective 
SB increases sharply upon freezing. At fields up to $\sim $400G this 
SB transition coincides with the FOT. At higher fields, however, we 
observe a remarkable behavior in which the SB transition follows a 
new line, labeled $T_{x}$ in Fig. 2, which resides above the FOT 
line and extends to elevated fields. Figure 3a shows this SB 
transition at 1000G. The sharp drop in $B_{ac}^{2f}$ at $T_{x}$ is 
very similar to that at $T_{FOT}$ at 300G. In addition, 
$B_{ac}^{2f}$  is still finite in the range $T_{d}<T<T_{x}$ as 
compared to vanishing $B_{ac}^{2f}$ at 300G. This means that the SB 
in phase B in Fig. 2 has an intermediate value between the strong 
barrier in phase E and a weak SB in phase C. Indeed, in the field 
range 400G to 750G the SB displays two transitions, exemplified 
below.

An alternative method to investigate the SB is by measuring the 
first harmonic $B_{ac}$ in the presence of a dc bias current. In the 
absence of a bias, vortices cross the sample in opposite directions 
on the positive and negative half-cycles of the ac current. 
Therefore, on average, each edge carries half of the current. Adding 
a dc bias breaks this symmetry.  Figure 3b shows an example of 
$B_{ac}$ at 650G due to 4mA ac current with and without bias current 
of $\pm $6.5mA, as measured by sensor 3 which is close to the left 
edge. At +6.5mA bias, vortices enter the sample only from the left. 
Therefore, at any time, the left barrier is larger than the right 
one and thus more than half of the current is flowing on the left 
edge. As a result a larger $B_{ac}$ (more negative), as compared to 
zero bias, is measured by sensor 3. For -6.5mA bias less than half 
of the current flows on the left edge and the signal decreases. The 
degree of splitting between the positive and negative bias curves 
reflects the strength of the SB. Above $T_{x}$ the SB is weak and 
the splitting is large. At $T_{x}$ the SB is enhanced significantly 
and the splitting is reduced by about a factor of two. An additional 
discontinuous drop in the splitting occurs at $T_{FOT}$ due to a 
further increase of the SB in the solid phase. At fields below 400G 
the two transitions merge into one FOT.

{\it Discussion}. Previous studies of the vortex matter phase 
diagram in BSCCO have identified three transition lines \cite 
{borisprl,boris-disorder,cubitt,lee} which meet at a multicritical 
point in Fig. 2: the first-order transition line $T_{FOT}$, the 
second peak line $B_{sp}$, and the upper depinning line $T_{d}$ 
(above $B_{sp}$). The corresponding three phases were interpreted as 
a highly disordered entangled vortex solid in phase A, one vortex 
fluid phase in B and C, and phases D and E as one quasi-lattice or 
Bragg-glass phase \cite{ertasnelson,gl-vinokur}.  Here we report two 
additional lines, the $T_{x}$ line, and the lower $T_{d}$ line 
separating the D and E phases (see also Ref. 22). We find that the 
vortices are mobile in phase E and immobile in phase D. Therefore 
one possible explanation is that the lower $T_{d}$ is depinning 
within the Bragg-glass phase \cite{ertasnelson,dewhurst}. An 
interesting observation is that this $T_{d}$ line drops almost 
vertically down from the multicritical point, and thus, 
alternatively, it may reflect a recently suggested phase transition 
line \cite{Horovitz}. Accordingly, phase E is a weakly disordered 
elastic Bragg glass whereas phase D is a strongly disordered 
Josephson glass \cite {Horovitz}.

We now discuss the $T_{x}$ line. Recent flux transformer studies 
\cite {sublimation} suggest that at high temperatures the FOT is a 
sublimation transition, and hence phase E is a lattice of vortex 
lines and phase C is a gas of pancakes. Our results here are 
consistent with this conclusion showing that phase E is the most 
ordered and coherent one, whereas phase C is highly uncorrelated. 
Furthermore, the SB behavior suggests the existence of an additional 
phase B which has a distinct structure with an intermediate degree 
of order. An important independent insight into the structure of 
phase B comes from a recent small angle neutron scattering study 
\cite {forganlt21+private}. The anisotropy of the BSCCO crystal used 
in these SANS experiments is very similar to ours, resulting in 
$T_{FOT}$ and $B_{sp}$ lines \cite{cubitt,forganlt21+private,kesrev} 
which are practically identical to those in Fig. 2. The earlier SANS 
data \cite{cubitt} indicated that Bragg peaks are present only in 
the D and E phases. However a more accurate study demonstrates that 
the line along which the Bragg peaks disappear follows the FOT line 
at high temperatures, but then shows an unexpected upturn and 
continues above the FOT in a ``concave'' form following our $T_{x}$ 
line \cite{forganlt21+private}. Furthermore, at fields above 
$B_{sp}$ the SANS data show a unique reentrant behavior in which 
weak but clearly resolvable Bragg peaks are present in the region 
between $T_{d}$ and $T_{x}$ in Fig. 2. For example, Bragg peaks are 
reported at 900G and 45K \cite{forganlt21+private}, indicated by a 
cross in Fig. 2. These results directly support our conclusion that 
phase B has a structure with intermediate order between that of 
phases E and C. Another important feature of phase B is gleaned from 
the study of very low doses of columnar defects 
\cite{boris-disorder}. It is found that the vortex matter loses its 
shear modulus along the $T_{FOT}$ line rather than along the $T_{x}$ 
line. Phase B therefore has zero or very low shear modulus.

Since Bragg peaks are observable at temperatures above the FOT, 
phase B cannot be an entangled vortex liquid.  One possibility is 
that this phase is a disentangled liquid of lines, as was proposed 
to exist in YBCO \cite{samoilov}, with some sort of hexatic order 
that results in weak Bragg peaks. Another possibility is the 
supersolid \cite{frey,dima}, decoupled solid 
\cite{glazkosh-Daemen,Horovitz}, or soft solid \cite{CarruzzoYu} 
phase, which consists of a rather well aligned stack of ordered 2D 
pancake layers. Such a phase is predicted to have a very low shear 
modulus due to vortex vacancies or interstitials, and is expected to 
display Bragg peaks. This interesting scenario implies that the 
$T_{FOT}$ is a first order-decoupling \cite 
{glazkosh-Daemen,Horovitz,doyle} or softening \cite{CarruzzoYu} 
transition rather than melting, $T_x$ is melting of the supersolid, 
and $B_{sp}$ is a disorder induced decoupling \cite{Horovitz} rather 
than a disorder induced solid entanglement 
\cite{ertasnelson,gl-vinokur}. Finally, as will be described 
elsewhere, the $T_d$ line displays some current and frequency 
dependence, whereas the $T_{FOT}$ and $T_x$ lines do not.  It 
therefore remains to be determined which of the lines in Fig. 2 
reflect thermodynamic phase transitions.

We thank V. Geshkenbein, A. Stern, B. Horovitz, M. Konczykowski, and 
R.  Doyle for valuable discussions. This work was supported by the 
Israel Ministry of Science and the Grant-in-Aid for Scientific 
Research from the Ministry of Education, Science, Sports and 
Culture, Japan, by the Israel Science Foundation, and by the MINERVA 
Foundation, Munich, Germany.


\newpage 

\ FIGURE CAPTIONS

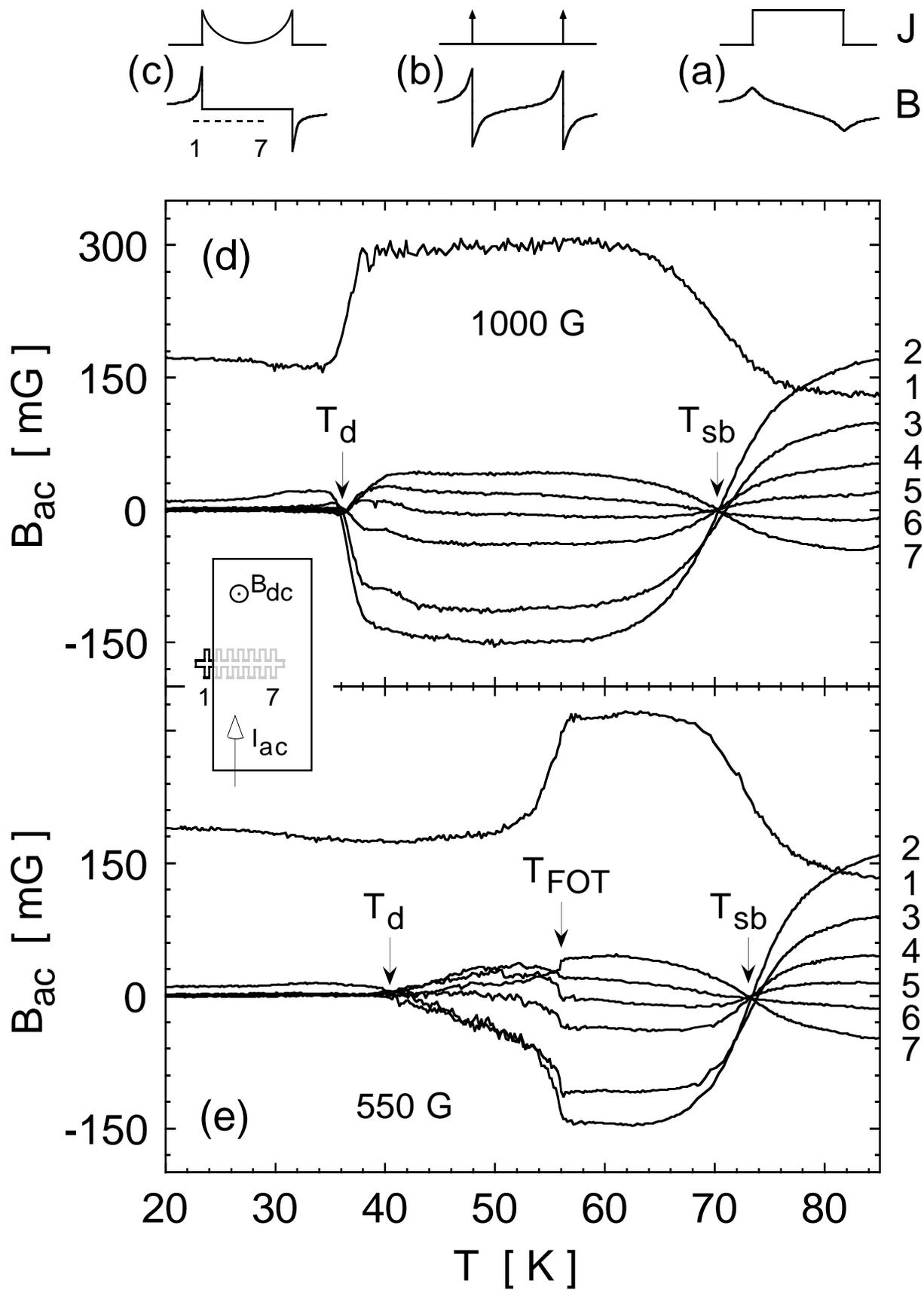
Fig. 1. Schematics of current distribution $J$, and corresponding 
self-field $B$, of the transport current for (a) uniform current 
flow, (b) surface barriers, and (c) bulk pinning, with sensor 
locations indicated. Self-field $B_{ac}$ induced by 4 mA ac current 
{\it vs}. $T$ in dc fields of 1000G (d) and 550G (e). Curves are 
labeled according to sensor number. {\it Inset:} Schematic top view 
of crystal attached to an array of seven Hall sensors.  Sensor 1 is 
outside the sample and the other six span more than half of the 
sample width.

Fig. 2. BSCCO vortex matter phase diagram. Vortices are immobile in 
phases A and D. Phase B has an intermediate ordered structure 
between the ordered lattice in phase E and a pancake gas in phase C. 
The $T_{x}$ line merges with the $T_{FOT}$ line at low fields. 
$T_{sb}$ indicates a crossover from uniform to SB dominated current 
flow.

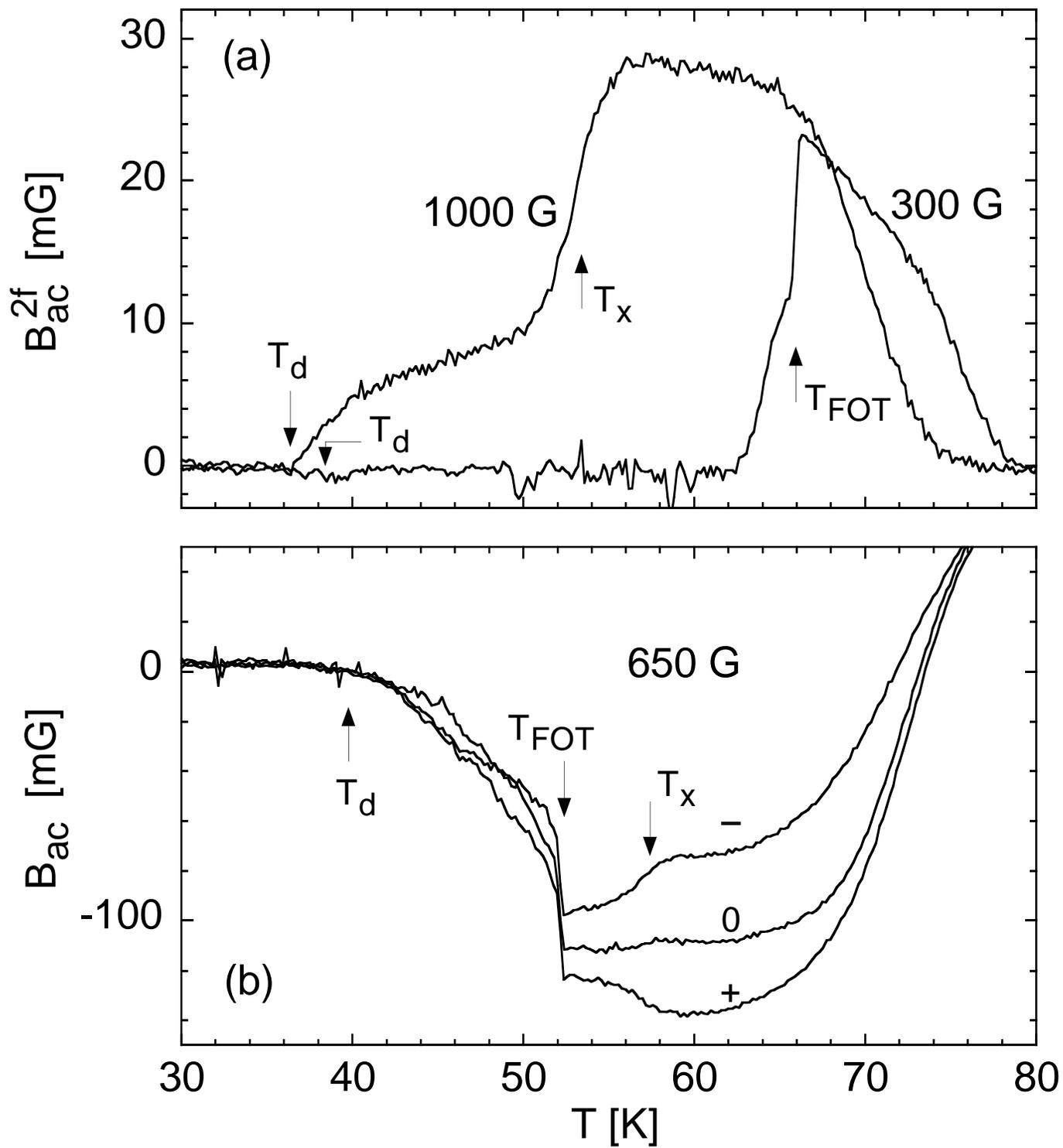
Fig. 3. (a) Second harmonic $B_{ac}^{2f}$ due to SB measured by 
sensor 3 at 10 mA ac current at 300 and 1000G. The SB is enhanced 
abruptly below $T_{FOT}$ and $T_{x}$, respectively. (b) First 
harmonic self-field $B_{ac}$ at 4 mA ac current and 650G without 
bias (0) and with positive (+) and negative (-) dc bias of 6.5 mA. 
The splitting between the + and - curves is reduced at $T_{x}$ due 
to enhancement of the SB. An additional drop in the splitting occurs 
at $T_{FOT}$ due to further enhancement of the SB.


\end{document}